\documentclass[11pt, oneside]{article}   	
\usepackage{geometry}                		
\usepackage{graphicx}				
\usepackage{amsmath}
\usepackage{xcolor}								

\usepackage[ruled,vlined]{algorithm2e}

\usepackage{natbib}		
\usepackage{amssymb}
\usepackage{placeins}
\usepackage{mathtools}

\title{Correcting spatial Gaussian process parameter and prediction variance estimation under informative sampling}
\author{Erin M. Schliep, Christopher K. Wikle \& Ranadeep Daw}

\begin{document}
\maketitle

\abstract{Informative sampling designs can impact spatial prediction, or kriging, in two important ways. 
First, the sampling design can bias spatial covariance parameter estimation, which in turn can bias spatial kriging estimates. Second, even with unbiased estimates of the spatial covariance parameters,
since the kriging variance is a function of the observation locations, these estimates will vary based on the sample and overestimate the population-based estimates. 
In this work, we develop a weighted composite likelihood approach to improve spatial covariance parameter estimation under informative sampling designs. Then, given these parameter estimates, we propose three approaches to quantify the effects of the sampling design on the variance estimates in spatial prediction. 
These results can be used to make informed decisions for population-based inference. 
We illustrate our approaches using a comprehensive simulation study. 
Then, we apply our methods to perform spatial prediction on nitrate concentration in wells located throughout central California.}


\section{Introduction}
An important objective in spatial analyses is to make predictions with estimates of uncertainty of one or more spatial processes at unobserved locations given data collected at a set of observed locations. 
In traditional geostatistics, the set of observed locations, e.g., the sample, is finite while the population, consisting of all the locations within a continuous field that could be observed or sampled, is infinite. 
In some spatial analyses, however, the population is also finite. 
For example, there are a finite number of groundwater wells across central California to access underground aquifers or, in the American Community Survey, each census tract consists of a finite number of households.
Even with a finite population, the entire population is rarely observed and the set of unobserved locations could be unknown.
Sampling designs, such as systematic designs, stratified designs, or spatially-balanced designs, aim to produce sample data that are representative of the population in order to infer population characteristics.
The sampling design for groundwater wells might aim to optimize spatial coverage across the region whereas the sampling design of the American Community Survey is stratified such that tracts with fewer households are sampled at a higher rate.
The sampling design has been shown to impact spatial prediction estimates at unobserved locations \citep{Gelfand2012, Pati2011, Diggle2010}.
Here, we study the effects of the sampling design in spatial Gaussian process parameter estimation, spatial prediction, and uncertainty estimation given a finite population.

We begin with a general definition of informative sampling designs. 
Let $Z(\mathbf{s})$ denote the process of interest.
Next, assume that $\mathcal{X}$ is the set of population locations and $\mathcal{X}_s$ is the set of sample locations.
\emph{Informative sampling} assumes that the distribution of the variable of interest, $Z$, conditional on the sample locations, $\mathcal{X}_s$, is not the same as that depending on the whole population, $\mathcal{X}$. 
Let $p(\mathbf{s})$ denote the sampling probability process. 
In survey statistics, $p(\mathbf{s})$ is considered known for all $\mathbf{s} \in \mathcal{X}_s$ and provides information regarding the sampling design. 
It is often referred to as the sample inclusion probability for location $\mathbf{s}$. 
Informative sampling designs in spatial statistics have been studied primarily in the context of geostatistics under the notion of \emph{preferential sampling}, as first introduced by \cite{Diggle2010}. 
Preferential sampling is defined such that the joint distribution $[Z,\mathcal{X}_s] \ne [Z][\mathcal{X}_s]$.
Let $\lambda_s(\mathbf{s})$ denote the location distribution function of $\mathcal{X}_s$. 
Under preferential sampling, a joint model is specified for $Z(\mathbf{s})$ and $\lambda_s(\mathbf{s})$ since the sample design is typically unknown.

Spatial prediction, or kriging, is often approached as a two stage procedure. 
At the first stage, spatial dependence in $Z(\mathbf{s})$ is modeled by computing the empirical semivariogram.
Under non-informative or non-preferential sampling, the empirical semivariogram ordinates are unbiased estimates of the theoretical semivariogram \citep{Zimmerman2010}. 
Scatter plots of the empirical and theoretical estimates can be used to suggest parametric semivariogram functions for the data. Customary methods for estimating the semivariogram parameters are ordinary least squares, weighted least squares, maximum likelihood, and restricted maximum likelihood \citep{Zimmerman1991}. 
More recently, composite likelihood methods for semivariogram parameter estimation were proposed \citep{Lele1997}. 

Given semivariogram parameter estimates obtained from any of these approaches, the second stage is to use kriging equations \citep{Cressie1993} to obtain prediction means and variances of the process at unobserved locations. 
Whereas the kriging mean is a function of the observed process, the kriging variance is only indirectly a function of the observed process through the semivariogram parameter estimates.
Specifically, given a semivariogram function and a set of parameter estimates, the kriging variance is a function of only the locations for which the data were observed and the locations where predictions are desired. 

Sampling designs can impact one or both of the stages of spatial kriging. 
\cite{Diggle2010} illustrate bias in variogram estimation under preferential sampling. 
Specifically, under non-preferential sampling designs that are either spatially random or clustered, empirical semivariogram estimates are shown to be unbiased. 
Under preferential sampling, the bias can be severe under either sampling design. 
This bias is the result of the reduction in the range of values of the response variable of interest, which, in turn, reduces the expected pairwise squared differences for a given spatial lag.  
For semivariogram estimation, a bias-corrected empirical semivarigram estimate has been suggested to account for and detect informative or preferential sampling where weights are defined as the inverse sampling intensity \citep{Rathbun}. 
This approach stems from the mark variogram estimator under preferential sampling \citep{HoStoyan2008}, where parameter estimates can be obtained using method of moments.

\cite{Gelfand2012} argue that the impact of preferential sampling is more important for prediction surfaces than parameter estimation. 
They illustrate remarkable differences between the spatial prediction surfaces when the observations are chosen randomly versus chosen preferentially. 
In geostatistical modeling, Bayesian approaches have been developed to account for preferential sampling \citep[e.g.,][]{Pati2011, Diggle2010, Gelfand2012}.
These approaches specify a spatial point process model (e.g., log-Gaussian Cox process) for the sample location distribution, $\lambda_s(\mathbf{s})$, such that the models for $Z(\mathbf{s})$ and $\lambda_s(\mathbf{s})$ have a shared spatial random effect to adjust for the spatial intensity of the sampling design in the process of interest.  
The focus of this specification is on the first order effects for improved inference and prediction rather than uncertainty estimation. 

\begin{figure}[h]
\begin{center}
\includegraphics[scale=.5]{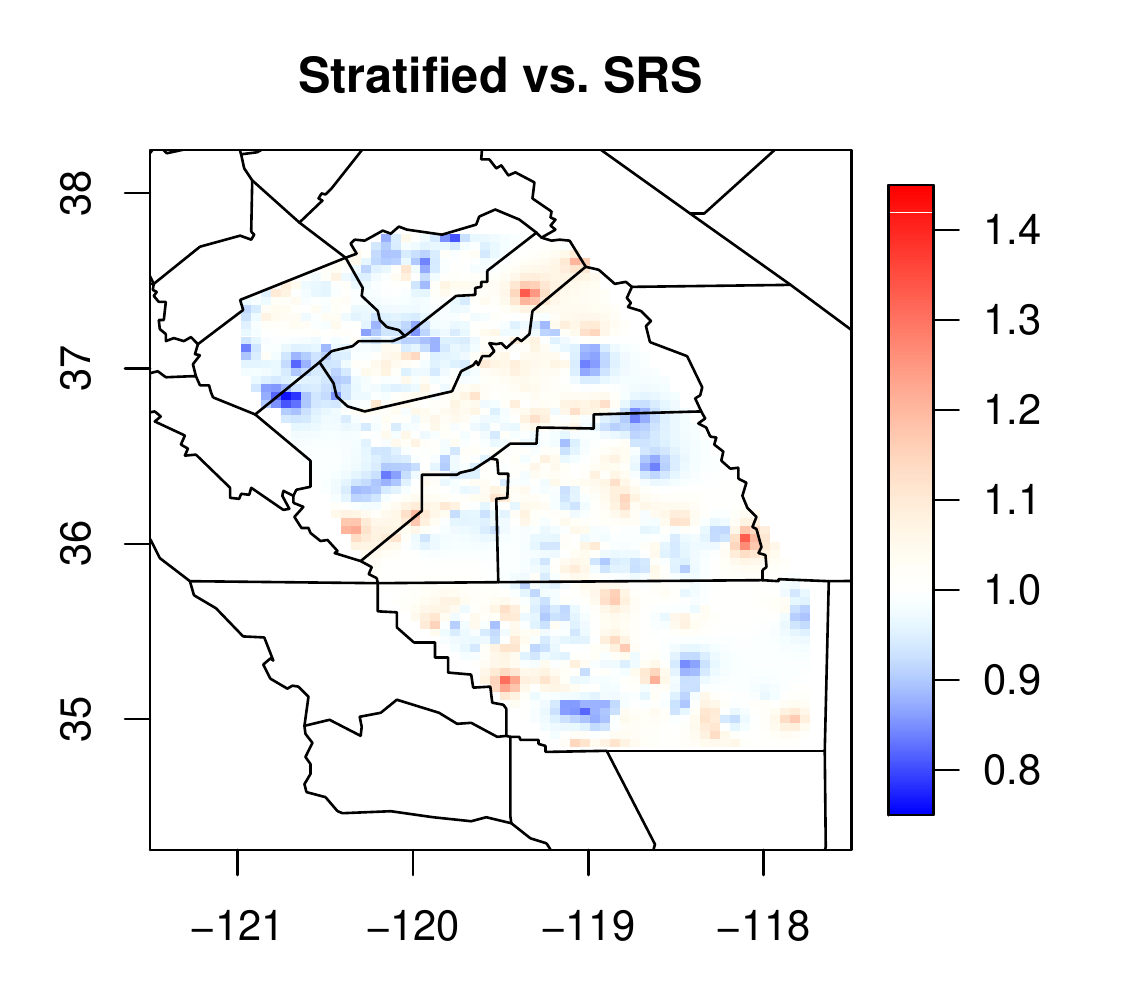}
\caption{Ratio of kriging variance estimates across the region under stratified sampling compared to simple random sample (SRS). \label{Mot}}
\end{center}
\end{figure}

Kriging variance estimates can vary dramatically across the spatial region based on different sample designs.
To illustrate, we turn to nitrate concentration data collected in wells in California. 
Note that these data are discussed and analyzed more thoroughly in Section \ref{sec:Wells} and similar data were analyzed in \cite{Chan2020}.
The population data consist of approximately 7000 wells in central California. 
The spatial covariance model is assumed fixed and known. 
We generate two sample realizations to illustrate the impact of sampling design on variance estimation. 
The two designs include a simple random sample and a spatially stratified sample. 
Given each set of sample locations, we obtain kriging variance estimates throughout the region. 
Figure \ref{Mot} shows the ratio of the kriging variance estimates from the stratified sample relative to the simple random sample. 
Regions in red (blue) indicate kriging variances that are greater (less than) those obtained under simple random sampling. 
These deviations in variance estimates vary across the region as well as vary between the different sampling designs. 
Understanding these impacts of sampling design on spatial kriging uncertainty is the focus of this work.

In survey data, informative sampling designs are accounted for in unit-level modeling and inference through the use of sampling weights, as done in the Horvitz-Thompson estimator for population inference \citep{Horvitz1952}.
A general approach for likelihood-based inference with survey data and informative sampling designs is the use of pseudo-likelihoods, where each unit's contribution to the likelihood is weighted by its survey weight \citep{Lumley2017}. 
We note that the term ``pseudo-likelihood" was first introduced by \cite{Besag1975} in the context of analyzing data having spatial dependence. 
His approach leveraged Markovian properties of auto-normal models and approximated the likelihood as the product of marginal densities. 
While the pseudo-likelihood is not the true likelihood function except in the case of independence, it is a computationally simpler method for parameter estimation than traditional likelihood estimation. 
See \cite{Parker2019} for a more detailed review of modern methods for unit-level modeling under informative sampling.

The contribution of this work is the development of approaches for incorporating sampling designs in (1) the semivariogram parameter estimation stage and (2) the kriging equations for variance estimation. 
We turn to the composite likelihood approach of \cite{Lele1997} for semivariogram estimation. 
To account for the sampling design, we define a weighted composite likelihood where the weights are either based on the survey weight or sampling intensity.
Our weighted composite likelihood approach provides a general framework for semivarioagram parameter estimation under informative and preferential sampling.
Given a semivariogram function and a set of parameter estimates, we then develop three schemes for obtaining kriging variance estimates to address the impacts of the sampling design.

The remainder of this paper is outlined as follows. 
In Section \ref{sec2}, we briefly review semivariogram estimation and then introduce our weighted composite likelihood approach to account for sampling designs. 
In Section \ref{sec3} we propose three approaches for obtaining kriging variance estimates under informative sampling designs.
We evaluate our approaches in the two stage spatial prediction procedure using a simulation study in Section \ref{sec:Sim}. 
In Section \ref{sec:Wells} we apply our approach using two common sampling designs to estimate nitrate concentrations in wells across central California. 
Lastly, we conclude with a discussion and directions for future work in Section \ref{sec:Disc}.

\section{Variogram parameter estimation}
\label{sec2}
Classical approaches for semivariogram estimation are method of moments, maximum likelihood (ML), and restricted maximum likelihood (REML). See \cite{Zimmerman1991} for a complete review and comparison of these approaches. 
Method of moments \citep{Matheron1963} is likely the most common approach to semivariogram estimation where parameter estimates are obtained by ordinary least squares or weighted least squares \citep{Cressie1993}. 
It benefits from being robust and requires no strong distributional assumptions about the process of interest. 
In addition, plotting the empirical semivariogram is a valuable exploratory tool for investigating spatial dependence as a function of distance. 
Unfortunately, the method of moments approach also has the significant drawback of requiring subjective specification of distance lags and lag tolerance regions. 
Maximum likelihood and restricted maximum likelihood approaches are consistent estimators and benefit from not having to make these specifications. 
Yet, they require a fully specified probabilistic model for the process and are often computationally costly by involving large matrix inversion. 

\cite{Lele1997} proposed composite likelihood methods as an alternative approach for semivariogram estimation. 
The composite likelihood (CL) approach is computationally and mathematically simpler and less reliant on strong distributional assumptions. 
Specifically, the benefits of the composite approach over the approaches previously mentioned are that it (i) eliminates subjectively specifying the distance lags and lag tolerance regions, (ii) eliminates needing to specify a probabilistic model, and (iii) is computationally efficient.
We begin with a brief review of the CL approach of \cite{Lele1997} and then extend to the weighted composite likelihood approach to account for sampling designs.

Let $Z(\mathbf{s})$ denote the spatial process of interest where $\mathbf{s}$ represents a spatial coordinate in the domain $\mathcal{D}$, often assumed to be two-dimensional. 
Under ordinary kriging \citep{Cressie1993}, it is assumed that ${E}(Z(\mathbf{s})) = \mu$. 
The semivariogram for the spatial process, $Z(\mathbf{s})$ is given by
$$\gamma(\mathbf{s}_i, \mathbf{s}_j) = \frac{1}{2} {var}(Z(\mathbf{s}_i) - Z(\mathbf{s}_j)).$$
Stationarity in the process implies  the semivariogram reduces to a function of the difference between the two locations, $\gamma(\mathbf{s}_i, \mathbf{s}_j) = \gamma(\mathbf{s}_i - \mathbf{s}_j)$. In that case, the covariance between $Z(\mathbf{s}_i)$ and $Z(\mathbf{s}_j)$ can be defined as 
\begin{equation}
C(Z(\mathbf{s}_i),Z(\mathbf{s}_j)) = C(\mathbf{s}_i - \mathbf{s}_j) = var(Z(\mathbf{s})) - \gamma(\mathbf{s}_i- \mathbf{s}_j).
\label{eq:1}
\end{equation}
Let $||\mathbf{s}_i - \mathbf{s}_j||$ denote Euclidean distance between $\mathbf{s}_i$ and $\mathbf{s}_j$.  When the semivariogram depends only on the distance between locations such that $\gamma(\mathbf{s}_i, \mathbf{s}_j) = \gamma(||\mathbf{s}_i - \mathbf{s}_j||)$, it is called isotropic. 
As such, we can write (\ref{eq:1}) as
\begin{equation}
C(Z(\mathbf{s}_i),Z(\mathbf{s}_j)) = C(||\mathbf{s}_i - \mathbf{s}_j||) = var(Z(\mathbf{s})) - \gamma(||\mathbf{s}_i- \mathbf{s}_j||).
\label{eq:2}
\end{equation}

For $\mathbf{s}_1, \dots, \mathbf{s}_n \in \mathcal{D}$, the set of contrasts is defined $\mathcal{V} = \{v_{ij}: i, j = 1, \dots, n\}$ where $v_{ij} = Z(\mathbf{s}_i) -Z(\mathbf{s}_j)$. 
Then, $$var(v_{ij}) = 2 \gamma(d_{ij}, \boldsymbol{\boldsymbol{\phi}}),$$ 
where $\boldsymbol{\phi}$ represents the parameters of the semivariogram and $d_{ij} =||\mathbf{s}_i - \mathbf{s}_j||$.
\cite{Lele1997} defines the composite likelihood as the product of the marginal densities of the contrasts. Namely, 
$$\mathcal{CL}(\boldsymbol{\phi}; \mathcal{V}) = \prod_{i=1}^{n-1}\prod_{j>i} f(v_{ij};\boldsymbol{\phi}).$$
where each component of the composite likelihood, $f(v_{ij};\boldsymbol{\phi})$ is a likelihood since $f(\cdot)$ is restricted to the class of valid density functions \citep{Lindsay1988}.

\cite{Curriero1999} show that the composite likelihood estimate is unbiased as long as $E(Z(\mathbf{s}_i) - Z(\mathbf{s}_j)) = 2\gamma(d_{ij},\boldsymbol{\phi})$. That is, the consistency of the estimator is not reliant on the correct specification of the marginal distribution of the contrasts, $v_{ij}$.
This is not true for maximum likelihood or restricted maximum likelihood approaches where misspecified joint distributions result in biased estimates.
Following \cite{Curriero1999} and for simplicity, we let $v_{ij} \sim N(0, 2\gamma(d_{ij};\boldsymbol{\phi}))$ such that
$$f(v_{ij};\boldsymbol{\phi}) = \frac{1}{\sqrt{2\pi} \sqrt{2\gamma(d_{ij};\boldsymbol{\phi})}}exp\left\{-\frac{(Z(\mathbf{s}_i) - Z(\mathbf{s}_j))^2}{4\gamma(d_{ij};\boldsymbol{\phi})}\right\}.$$
Then, the log-composite likelihood is written 
\begin{equation}\text{log}~ \mathcal{CL}(\boldsymbol{\phi};\mathcal{V}) \propto \sum_{i=1}^{n-1} \sum_{j>i} \left\{-\frac{(Z(\mathbf{s}_i) - Z(\mathbf{s}_j))^2}{2\gamma(d_{ij};\boldsymbol{\phi})} - \text{log}(\gamma(d_{ij};\boldsymbol{\phi})) \right\}
\label{eq:cl}
\end{equation}
and composite likelihood estimates of the parameters of $\boldsymbol{\phi}$ are those that maximize (\ref{eq:cl}). 

Similar to restricted maximum likelihood, the composite likelihood approach eliminates the nuisance parameter $\mu$ by considering the contrasts $v_{ij} = Z(\mathbf{s}_i) - Z(\mathbf{s}_j)$. That is, the approach is based solely on the semivariogram parameters of interest.
The composite likelihood approach and objective function in (\ref{eq:cl}) is computationally simpler than the maximum likelihood and restricted maximum likelihood objective functions as it requires no matrix inversion.

\cite{Bevilacqua2012} developed an extension to this composite likelihood approach for space-time covariance functions. They proposed a weighted version of the composite likelihood approach where cutoff weights, $w_{ij} \geq 0$, are based on distance in space and time. That is, pairs of observations that are far apart in time and space are down-weighted such that they contribute less to the estimation of the semivariogram parameters. 
The log weighted composite likelihood is written 
\begin{equation}\text{log}~ \mathcal{WCL}(\boldsymbol{\phi};\mathcal{V}) \propto \sum_{i=1}^{n-1} \sum_{j>i} w_{ij} \left\{-\frac{(Z(\mathbf{s}_i) - Z(\mathbf{s}_j))^2}{2\gamma(d_{ij};\boldsymbol{\phi})} - \text{log}(\gamma(d_{ij};\boldsymbol{\phi})) \right\}.
\label{eq:wcl}
\end{equation}
The weighted composite likelihood has been shown to have both computational advantages as well as improved statistical efficiency.

Here, we propose the weighted composite likelihood approach to semivariogram estimation as a way to incorporate sampling designs (e.g., survey weights, preferential sampling) into the likelihood to improve parameter estimation. 
We offer two different weighting schemes, the first stemming from the survey sampling literature and the second from geostatistics under preferential sampling.

\subsection{Survey design weighted composite likelihood estimation}
Our first approach is to use survey weights in the weighted composite likelihood in (\ref{eq:wcl}) where we define $w_{ij}$ as the weight assigned to the pair of observations at locations $\mathbf{s}_i$ and $\mathbf{s}_j$.
In the survey literature, Binder (1983) and Skinner (1989) proposed \emph{pseudo-likelihood analysis} as an approach for incorporating survey weights into the likelihood to get design-consistent parameter estimates. 
For example, the pseudo-log likelihood can be written 
$$\sum_{i \in \mathcal{S}} w_i ~\text{log} f(Z(\mathbf{s}_i); \boldsymbol{\theta})$$
where $f(Z(\mathbf{s}_i); \boldsymbol{\theta})$ is the specified density function of the data having parameters $\boldsymbol{\theta}$, $\mathcal{S}$ denotes the set of sampled observations from some known population, and $w_i = \frac{1}{p(\mathbf{s}_i)}$ is the inverse sampling probability associated with observation $Z(\mathbf{s}_i)$. 
Here, we propose incorporating survey weights into the weighted composite likelihood. 
Assuming sampling is done independently, we might define the weights to be the product of the inverse  sampling probabilities from the survey design for observation $i$ and $j$. That is, $w_{ij} = \frac{1}{p(\mathbf{s}_i)}\frac{1}{p(\mathbf{s}_j)}$. As such, our first log weighted composite likelihood is
\begin{equation}\text{log}~ \mathcal{WCL}_1(\boldsymbol{\phi};\mathcal{V}) \propto \sum_{i=1}^{n-1} \sum_{j>i} \frac{1}{p(\mathbf{s}_i)}\frac{1}{p(\mathbf{s}_j)} \left\{-\frac{(Z(\mathbf{s}_i) - Z(\mathbf{s}_j))^2}{2\gamma(d_{ij};\boldsymbol{\phi})} - \text{log}(\gamma(d_{ij};\boldsymbol{\phi}) \right\}.
\label{eq:wcl1}
\end{equation} 
Parameter estimates of $\boldsymbol{\phi}$ can be obtained by maximizing (\ref{eq:wcl1}).

\subsection{Sampling design weighted composite likelihood estimation}

The second approach stems from geostatistics under preferential sampling. The bias-corrected estimator for the mark semivariogram under preferential sampling \citep{HoStoyan2008, Rathbun} is 
\begin{equation*}
\hat{\gamma}(h) = \frac{\sum_{i \neq j} k\left(\frac{h - d_{ij}}{b}\right) \lambda_s^{-1}(\mathbf{s}_i) \lambda_s^{-1}(\mathbf{s}_j) (Z(\mathbf{s}_i) - Z(\mathbf{s}_j))^2}{\sum_{i \neq j} k\left(\frac{h - d_{ij}}{b}\right) \lambda_s^{-1}(\mathbf{s}_i) \lambda_s^{-1}(\mathbf{s}_j)}
\end{equation*}
where $k(\cdot)$ is a kernel density function with bandwidth $b$, and $\lambda_s(\mathbf{s})$ is the intensity function of the sample locations.
This is a weighted version of the method of moments estimator where the weights are a function of the 
intensity, $\lambda_s(\mathbf{s})$, and the kernel density function specification corresponds to the subjectivity in the lag tolerance regions discussed above.

For our second weighted composite likelihood, we propose using the inverse intensity estimates as our weights. That is, 
\begin{equation}\text{log}~ \mathcal{WCL}_2(\boldsymbol{\phi};\mathcal{V}) \propto \sum_{i=1}^{n-1} \sum_{j>i} \frac{1}{\lambda_s(\mathbf{s}_i)}\frac{1}{\lambda_s(\mathbf{s}_j)} \left\{-\frac{(Z(\mathbf{s}_i) - Z(\mathbf{s}_j))^2}{2\gamma(d_{ij};\boldsymbol{\phi})} - \text{log}(\gamma(d_{ij};\boldsymbol{\phi}) \right\}.
\label{eq:wcl2}
\end{equation} 
To estimate the semivariogram parameters, we must first estimate the intensity function $\lambda_s(\mathbf{s})$ for the spatial domain, $\mathcal{D}$.
Estimates of $\boldsymbol{\phi}$ can be obtained by plugging in estimates of $\lambda_s(\mathbf{s})$ and maximizing (\ref{eq:wcl2}). 
Note that while our focus is on informative sampling designs where the sampling rate is known, this approach can also be used when the sampling design is unknown. 
While this approach does not take uncertainty quantification associated with estimates of $\lambda_s(\mathbf{s})$ into account, simulations showed minimal sensitivity to using the estimates rather than the true intensity. 
Methods for incorporating uncertainty in the estimation of $\lambda_s(\mathbf{s})$ into (\ref{eq:wcl2}) is the subject of future research. 

\section{Kriging variance estimates}
\label{sec3}
Next, we turn to spatial prediction given variogram parameter estimates.
We propose three approaches for quantifying the effects of the sampling design on uncertainty estimation in spatial prediction for finite populations.
Under each approach, \emph{the goal is to obtain an estimate of the kriging variance had the population been observed}. 
We study the effects of the sampling design by comparing the kriging estimates obtained from the sample data to that which would have been obtained had we observed the entire population.
The sampling design is assumed known, however, we consider both the case when the unobserved locations within the population are known and unknown. 

We begin by reviewing ordinary kriging. 
Let $\mathbf{s}^*$ denote an unobserved location for which we would like to predict the spatial process $Z(\mathbf{s})$. That is, we want to predict $Z(\mathbf{s}^*)$ given observations $\mathbf{Z} = (Z(\mathbf{s}_1), \dots, Z(\mathbf{s}_n))'$ at population locations $\mathbf{s}_1, \dots, \mathbf{s}_n \in \mathcal{X}$. 
Let $\mathbf{c}^*$ denote the length $n$ cross-covariance vector with $i$th element $\text{cov}(Z(\mathbf{s}^*), Z(\mathbf{s}_i))$.  
Similarly, let $\mathbf{C}$ denote the $n \times n$ covariance matrix between all $Z(\mathbf{s}_i)$ and $Z(\mathbf{s}_j)$ for $\mathbf{s}_i, \mathbf{s}_j \in \mathcal{X}$. 
As shown in \cite{Cressie1993}, the generalized least squares estimator of $\mu$ is
$$\hat{\mu} =  \frac{\mathbf{1}_n' \mathbf{C}^{-1}\mathbf{Z}}{\mathbf{1}_n' \mathbf{C}^{-1}\mathbf{1}_n},$$
where $\mathbf{1}_n$ is a length $n$ vector of 1s.
In addition, the ordinary kriging predictor is 
\begin{equation}
\begin{split}
\hat{Z}_{ok}(\mathbf{s}^*) &= \left(\mathbf{c}^*+ \mathbf{1}_n \left(\frac{(1-\mathbf{1}'_n \mathbf{C}^{-1}\mathbf{c}^*)}{\mathbf{1}_n' \mathbf{C}^{-1}\mathbf{1}_n}\right)\right)'\mathbf{C}^{-1} \mathbf{Z}\\
&=\hat{\mu} + \mathbf{c}^{*'}\mathbf{C}^{-1}(\mathbf{Z} - \hat{\mu}\mathbf{1}_n).
\end{split}
\end{equation}

Here, we focus on the ordinary kriging variance, which we denote, $\sigma^2_{ok}(\mathbf{s}^*)$. 
Given $\mathbf{Z}$, the ordinary kriging variance is 
\begin{equation}
\sigma^2_{ok}(\mathbf{s}^*) = \text{var}(Z(\mathbf{s}^*)) - \mathbf{c}^{*'}\mathbf{C}^{-1}\mathbf{c}^*+ \frac{(1-\mathbf{1}'_n \mathbf{C}^{-1}\mathbf{c}^*)^2}{\mathbf{1}_n' \mathbf{C}^{-1}\mathbf{1}_n}.
\label{eq:OK1}
\end{equation}
Note that the first two terms are the simple kriging variance and the last term is the penalty for estimating $\mu$ \citep{Cressie1993}. The ordinary kriging variance is lower for locations that are close to observation locations. The variance increases as distance increases between the prediction location and the observations. 

The kriging variance estimate of $Z(\mathbf{s}^*)$ in (\ref{eq:OK1}) assumes the population set of all $n$ locations is observed. 
The kriging variance is minimized when the entire population is observed. That is, for a sample of size $m$ where $m < n$, the kriging variance is greater than that based on all $n$ locations.
Without loss of generality, let $\mathcal{X}_s = \{\mathbf{s}_1, \dots, \mathbf{s}_m\}$ denote the set of $m$ observed locations in the sample, and the remaining $\mathbf{s}_{m+1} \dots, \mathbf{s}_n$ denote the unobserved locations in the population. 
Dropping the dependence on $\mathbf{s}$ for ease of notation, we partition $\mathbf{Z} = (\mathbf{Z}_o, \mathbf{Z}_u)$ into the set of $m$ observed variables, $\mathbf{Z}_o$, and $n-m$ unobserved variables, $\mathbf{Z}_u$. 
We partition $\mathbf{c}^* = (\mathbf{c}^*_o, \mathbf{c}^*_u)$ into the cross-covariances between $Z(\mathbf{s}^*)$ and the variables at the observed set of locations and those unobserved. 
Similarly, let
$$\mathbf{C} = \left(\begin{array}{cc} \mathbf{C}_{oo} & \mathbf{C}_{ou}\\ \mathbf{C}_{ou}' & \mathbf{C}_{uu} \end{array}\right)$$
denote the covariance matrix between all $Z(\mathbf{s}_i)$ and $Z(\mathbf{s}_j)$ for $\mathbf{s}_i, \mathbf{s}_j \in \mathcal{X}$, which again we partition into the covariances between the observed set of locations and those unobserved. 
Now, the ordinary kriging predictor and variance estimate given only the observed locations are
\begin{equation}
\begin{split}
\hat{Z}_{ok,o}(\mathbf{s}^*) &=\hat{\mu}_o + \mathbf{c}^{*'}_o\mathbf{C}^{-1}_{oo}(\mathbf{Z}_o - \hat{\mu}_o\mathbf{1}_o)\\
\hat{\mu}_o &=  \frac{\mathbf{1}'_m \mathbf{C}^{-1}_{oo}\mathbf{Z}_o}{\mathbf{1}'_m \mathbf{C}^{-1}_{oo}\mathbf{1}_m}
\end{split}
\end{equation}
and 
\begin{equation}
\sigma^2_{ok,o}(\mathbf{s}^*) = \text{var}(Z(\mathbf{s}^*)) - \mathbf{c}^{*'}_{o}\mathbf{C}^{-1}_{oo}\mathbf{c}^*_{o} + \frac{(1-\mathbf{1}'_m \mathbf{C}^{-1}_{oo}\mathbf{c}^*_{o})^2}{\mathbf{1}'_m \mathbf{C}^{-1}_{oo}\mathbf{1}_m}.
\label{eq:OK2}
\end{equation}
Given a known covariance function and parameters, $\sigma^2_{ok,o} = \sigma^2_{ok}$ only when $m=n$ such that all of the locations in the population are in the sample.

If the unsampled locations $\mathbf{s}_{m+1} \dots, \mathbf{s}_n$ are known, the first and preferred approach is to use them when computing the kriging variance estimate. This approach is optimal in the sense that the kriging variance estimates will not be impacted by the sampling design since (\ref{eq:OK1}) can be computed directly given $\mathbf{s}_1, \dots, \mathbf{s}_n$. The other two approaches, which we detail below, assume that the unsampled locations are unknown. As mentioned above, the difference between the kriging variance computed using the sample and population decreases as sample size increases and the proximity of the sample locations to the prediction location decreases. Therefore, our two approaches aim to adjust the distance (scaling approach) between the observations and the prediction location and the number (simulation approach) of sampled locations.


\subsection{Scaling approach}
\label{sec:ScaleApp}

The scaling approach aims to adjust the distance in the spatial covariance function to account for the sample design. 
Assuming a stationary and isotropic covariance function, the scaling is done on the distance between the two locations. Alternatively, it can be thought of as a scaling of the spatial range of the covariance function.
The approach is motivated by the $K$-function \citep{Ripley1977} and $G$ function \citep{Baddeley2015}, measures commonly used in spatial point processes to detect deviations from complete spatial randomness (CSR).
Recall that, given lag distance $h$, $K(h)$ is defined as the expected number of events within $h$ of an event in the point pattern, divided by the average intensity of the point pattern. 
Alternatively, $G(h)$ defines the distribution of the distances from an event to its nearest event.

Let the random variable $N_r(\mathbf{s})$ denote the number of points within radius $r$ of the location $\mathbf{s}$. 
Assuming a homogenous point process (CSR) with intensity $\lambda(\mathbf{s}) = \lambda$, 
$$E(N_r(\mathbf{s})) = \pi r^2 \lambda.$$
If we set this equal to some value $v>0$, we can solve for $r$ and obtain
$$r = \sqrt{\frac{v}{\pi \lambda}}.$$
Thus, we expect $v$ points within a neighborhood with radius $\sqrt{\frac{v}{\pi \lambda}}$
around $\mathbf{s}$. 
Additionally, under complete spatial randomness, $G(h) = 1-\exp\{-\lambda \pi h^2\}$. 
Using these results, we outline the scaling approach under both simple random sampling and informative sampling.

\subsubsection{Simple random sampling}
Assume the sampling design is a simple random sample (SRS) with known rate $p$ where $0 < p < 1$. 
The intensity of the point pattern of the sample data is $\lambda p$. 
Let the random variable $\widetilde{N}_{\widetilde{r}}(\mathbf{s})$ denote the number of points in the sample within radius $\widetilde{r}$ of $\mathbf{s}$. 
We compute the expected value of $\widetilde{N}_{\widetilde{r}}(\mathbf{s})$ as
$$E(\widetilde{N}_{\widetilde{r}}(\mathbf{s})) = \pi \widetilde{r}^2 \lambda p.$$
That is, the expected number of points in the sample within radius $\widetilde{r}$ of $\mathbf{s}$ is a function of the sampling rate $p$. 
If we set $E(N_r(\mathbf{s})) = E(\widetilde{N}_{\widetilde{r}}(\mathbf{s}))$ and solve for ${r}$, we get
$$r = \sqrt{p}~\widetilde{r}$$
suggesting a distance scaling of ${\sqrt{p}}$ to relate the sample to the population.  
Under SRS, we can derive the analogous $G$-function using the sample data $G(\widetilde{h}) = 1-\exp\{-\lambda p \pi \widetilde{h}^2\}$. 
Setting $G(h) = G(\widetilde{h})$ and solving for $h$ results in 
$$h = \sqrt{p}~\widetilde{h}.$$ 
Again, this suggests that the distance from an event to its nearest event in the population is $\sqrt{p}$ times the distance from an event to its nearest event in the sample. 
Based on these two measures, we propose incorporating $\sqrt{p}$ distance scaling into the covariance function when doing spatial kriging of $Z(\mathbf{s})$ using sample data. 

Assume a stationary and isotropic covariance function, $ C(Z(\mathbf{s}_i), Z(\mathbf{s}_j)) = C(||\mathbf{s}_i - \mathbf{s}_j||; \boldsymbol{\phi})$.
Letting $C_p(||\mathbf{s}_i - \mathbf{s}_j||; \boldsymbol{\phi})$ denote the covariance function given the sampling rate $p$, we propose
\begin{equation}
C_p(||\mathbf{s}_i - \mathbf{s}_j||; \boldsymbol{\phi}) = C(\sqrt{p}||\mathbf{s}_i - \mathbf{s}_j||; \boldsymbol{\phi}).
\label{eq:krigV}
\end{equation}
That is, given the sample data, the covariance between $Z(\mathbf{s}_i)$ and $Z(\mathbf{s}_j)$ is a function of the distance between the two locations scaled by $\sqrt{p}$. 
To obtain the kriging variance of $Z(\mathbf{s}^*)$ using the sample data, we propose using the scale covariance function in (\ref{eq:krigV}) in place of the original covariance function in (\ref{eq:OK2}). 
%

\subsubsection{Informative sampling}
The scaling approach can also be employed under informative sampling. 
Let the sampling rate now vary as a function of location, where $p(\mathbf{s})$ is known for all $\mathbf{s} \in \mathcal{X}_s$. 
Recall that our goal is to obtain the kriging variance at location $\mathbf{s}^*$. 
In order to use the scaling approach, we must first estimate $p(\mathbf{s}^*)$ unless it is known. 
Assuming there exists spatial dependence in the sampling rate, we suggest using a kernel based smoother of $p(\mathbf{s})$ to estimate $p(\mathbf{s}^*)$.
Other model-based spatial prediction approaches, such as kriging $p(\mathbf{s})$ on the log scale, are also available.
In our simulation and application below, we use a Gaussian kernel smoother to obtain estimates of $p(\mathbf{s}^*)$.
To obtain the kriging variance of $Z(\mathbf{s}^*)$, we replace the covariance function in (\ref{eq:OK2}) with $C_{p(\mathbf{s}^*)}(||\mathbf{s}_i - \mathbf{s}_j||; \boldsymbol{\phi})$. 
Algorithm \ref{Alg1} outlines the scaling approach for estimating the kriging variance under both informative and non-informative sampling. 


\begin{algorithm}[]
\SetAlgoLined
\caption{Scaling approach for estimating the kriging variance of $Z(\mathbf{s}^*)$}
Given a set of sample locations $\mathbf{s}_1, \dots, \mathbf{s}_m$ and a stationary and isotropic covariance function $C(Z(\mathbf{s}_i),Z(\mathbf{s}_j)) =C(||\mathbf{s}_i - \mathbf{s}_j||; \boldsymbol{\phi})$,
\begin{enumerate} 
\item Obtain an estimate of $p(\mathbf{s}^*)$\\
{\bf if} simple random sample with probability $p$\\
~~~~~$p(\mathbf{s}^*) = p$\\
{\bf else} given values $p(\mathbf{s}_1) \dots p(\mathbf{s}_m)$, use a kernel smoother to obtain an estimate of $p(\mathbf{s}^*)$\\
\item Using the $m$ sample locations:
\begin{enumerate}
\item Define the $m \times m$ matrix, $\mathbf{C}_{oo}$: for all $i,j = 1, \dots, m$, $C(Z(\mathbf{s}_i),Z(\mathbf{s}_j)) =C(\sqrt{p(\mathbf{s}^*)}||\mathbf{s}_i - \mathbf{s}_j||; \boldsymbol{\phi})$
\item Define the length $m$ vector, $\mathbf{c}_o^*$: for all $i =1, \dots, m$, $C(Z(\mathbf{s}_i),Z(\mathbf{s}^*)) = C(\sqrt{p(\mathbf{s}^*)}||\mathbf{s}_i - \mathbf{s}^*||; \boldsymbol{\phi})$
\end{enumerate}
\item Compute the kriging variance of $Z(\mathbf{s}^*)$ according to (\ref{eq:OK2}).
\end{enumerate} 
\label{Alg1}
\end{algorithm}

\subsection{Simulation approach}
\label{sec:SimApp}
As discussed above, the kriging variance estimate at the prediction location is a function of the covariance parameters and the observed locations. 
Our third approach for estimating the kriging variance is simulation based, where we randomly locate unsampled locations to mimic the population in order to mitigate the variance inflation that results from the sample size being less than the population. 

Recall that in the survey designs considered here, we assume the total population size $n$ is known, but the unsampled locations are unknown. Under either informative or non-informative sampling, the simulation approach requires randomly locating $n-m$ events, which we will refer to as pseudo-observation locations, to be used when computing the kriging variance. That is, we generate pseudo-observation locations $\widetilde{\mathbf{s}}_{m+1}, \dots, \widetilde{\mathbf{s}}_n$ and then treat them as observed locations when computing the kriging variance in (\ref{eq:OK2}). 
For a stationary and isotropic covariance function, the closer the collection of pseudo-observation locations are to the set of the unsampled locations in terms of the distribution of their distances to the prediction location, the more accurate the simulation approach will be to estimating the population-based kriging variance.

\subsubsection{Simple random sampling}
Under simple random sampling, the point pattern of sample locations offers a thinned representation of the population point pattern. 
In fact, independently thinning a spatial point pattern by randomly assigning each point to either the training or test set is a common approach for model validation as it preserves the properties of the point process \citep{Leininger2017}.
As such, with a non-informative sampling design, the set of sampled locations can be used to approximate the intensity function of the population, which can be used to generate the pseudo-observation locations. 
For example, we might use kernel density estimation to obtain an estimate of the density surface, and then randomly locate the $n-m$ pseudo-observations independently across the spatial region using rejection sampling \citep{Gelfand2018}. 
That is, we randomly sample locations within the domain and retain them independently with probability proportional to the density estimate at the location. This procedure is repeated until $n-m$ pseudo-observation locations are collected.
Whereas the population point process might exhibit interaction between points, as in Gibbs processes with attractive (clustering) or repulsive (inhibition) interactions, we found through simulation that this second order structure can be captured well by the first order intensity for the purposes of locating pseudo-observations (results not shown).

\subsubsection{Informative sampling}
Under an informative sampling design, we must account for the fact that our kernel density estimate obtained from the point pattern of sampled locations is a biased representation of the population. 
That is, $\lambda_s(\mathbf{s}) = p(\mathbf{s}) \lambda(\mathbf{s})$ where $\lambda(\mathbf{s})$ is the population location distribution.
Therefore, in the accept-reject step of the procedure above, the probability of retaining the randomly sampled point as a pseudo-observation location is scaled by the reciprocal of the sample rate at the location. 
Note that this, again, requires spatial estimation (e.g., kernel smoothing, kriging) of the sample rates discussed in the scaling approach in Section \ref{sec:ScaleApp} to every proposed pseudo-observation location. 
Algorithm \ref{Alg2} outlines the scaling approach for estimating the kriging variance under both informative and non-informative sampling.

\begin{algorithm}[]
\SetAlgoLined
\caption{Simulation approach for estimating the kriging variance of $Z(\mathbf{s}^*)$}
Given a set of sample locations $\mathbf{s}_1, \dots, \mathbf{s}_m$, and a stationary and isotropic covariance function $C(Z(\mathbf{s}_i),Z(\mathbf{s}_j)) =C(||\mathbf{s}_i - \mathbf{s}_j||; \boldsymbol{\phi})$
\begin{enumerate} 
\item Obtain an estimate of the intensity function $\lambda(\mathbf{s})$ across the spatial domain, $\mathcal{D}$ \\

\item Randomly sample candidate $\mathbf{s}_c \in \mathcal{D}$\\
{\bf if} simple random sample with probability $p$\\
~~~~~retain $\mathbf{s}_c$ as pseudo-observation location with probability $\frac{\lambda(\mathbf{s}_c)}{max_{\mathbf{s}\in \mathcal{D}}(\lambda(\mathbf{s}))}$\\
{\bf else} given known values $p(\mathbf{s}_1) \dots p(\mathbf{s}_m)$, use a kernel smoother to obtain an estimate of $p(\mathbf{s}_c)$\\

~~~~~retain $\mathbf{s}_c$ as pseudo-observation location with probability $\frac{\lambda(\mathbf{s}_c)/p(\mathbf{s}_c)}{max_{\mathbf{s}\in \mathcal{D}}\lambda(\mathbf{s})/p(\mathbf{s})}$
\item Repeat step 2 until $n-m$ pseudo-observations are retained
\item Using all $m$ sample locations and $n-m$ pseudo-observation locations:
\begin{enumerate}
\item Define the $n \times n$ matrix, $\mathbf{C}$: For all $i,j = 1, \dots, m, m+1, \dots, n$, $C(Z(\mathbf{s}_i),Z(\mathbf{s}_j)) =C(||\mathbf{s}_i - \mathbf{s}_j||; \boldsymbol{\phi})$
\item Define the length $n$ vector, $\mathbf{c}^*$: For all $i =1, \dots, m, m+1, \dots, n$, $C(Z(\mathbf{s}_i),Z(\mathbf{s}^*)) = C(||\mathbf{s}_i - \mathbf{s}^*||; \boldsymbol{\phi})$
\end{enumerate}
\item Compute the kriging variance of $Z(\mathbf{s}^*)$ according to (\ref{eq:OK1}).
\end{enumerate} 
\label{Alg2}
\end{algorithm}

\section{Simulation study}
\label{sec:Sim}
The objective of our simulation study is to assess the impact of varying sampling designs on kriging estimates. 
The simulation procedure consists of the following steps: (i) simulate the population data as well as prediction locations, (ii) sample from the population under a specified sample design, (iii) obtain variogram parameter estimates, and (iv) obtain kriging estimates at the prediction locations.
Each of these steps is described below. We compare each of the approaches for parameter estimation and kriging variance proposed in Sections 2 and 3.

\subsection{Simulating the population}
We simulate data within the unit square using the following model specifications and parameterizations. Let $W(\mathbf{s})$ and $U(\mathbf{s})$ for $\mathbf{s} \in [0,1] \times [0,1]$ be two independent mean 0 Gaussian processes with stationary and isotropic covariance functions. For both processes, we assume an exponential covariance function with spatial variance $\sigma^2 = 0.4$, and range parameter $\phi = 0.1$.
The response variable of interest, $Z(\mathbf{s})$, is defined as
\begin{equation}
Z(\mathbf{s}) = \mu + W(\mathbf{s}) + \epsilon(\mathbf{s})
\label{eq:Z}
\end{equation}
where $\epsilon(\mathbf{s})$ represents independent measurement error. We assume $\mu=1$ and $\epsilon(\mathbf{s}) \stackrel{iid}{\sim}N(0, \tau^2$) where $\tau^2 = 0.2$. 

The set of population locations are simulated under two different scenarios. The first assumes that the location distribution is independent of the process of interest whereas the second assumes the location distribution and process of interest to be dependent. 
Let $\lambda_1(\mathbf{s})$ and $\lambda_2(\mathbf{s})$ denote the intensity functions assumed to be generating the population locations. We assume log-Gaussian Cox processes (LGCPs) for both $\lambda_1(\mathbf{s})$ and $\lambda_2(\mathbf{s})$. Thus, conditional on the intensity, the event locations are independent. Define
$$\lambda_1(\mathbf{s}) = \exp(\beta U(\mathbf{s}))$$
and 
$$\lambda_2(\mathbf{s}) = \exp(\beta W(\mathbf{s})).$$
In our simulations, $\beta = 1$.

We simulate from each LGCP to obtain the two sets of population locations, $\mathcal{X}_1$, and $\mathcal{X}_2$. Our simulation resulted in population sizes of $n_1 = 4365$ and $n_2 = 4332$. For each population, we simulate a realization of the process of interest $Z(\mathbf{s})$ according to (\ref{eq:Z}). Let $\mathbf{Z}_1$ and $\mathbf{Z}_2$ denote the observations of populations 1 and 2, respectively.
Lastly, we simulate $100$ prediction locations independently of $\mathcal{X}_1$ and $\mathcal{X}_2$ from a uniform distribution over the domain for which we are interested in obtaining kriging estimates.

For each population, estimates of the exponential variogram parameters were obtained by maximizing the composite likelihood. Using these variogram parameter estimates, kriging means and variances were obtained for the set of prediction locations. The parameter estimates, means, and variances obtained using the population data are used as the metric of comparison for each of the sampling designs and estimation procedures described below.

\subsection{Sampling designs}

We propose three different sampling designs (a), (b), and (c), for each population with varying degrees of being preferential and informative. For population $i =1, 2$, let $p_{ia}(\mathbf{s})$, $p_{ib}(\mathbf{s})$, and $p_{ic}(\mathbf{s})$ denote the sampling rate processes for the three sampling designs. Define the sampling location distribution for sampling design (a) and population 1 as, $\lambda_{1a}(\mathbf{s}) = p_{1a}(\mathbf{s})\lambda_1(\mathbf{s})$. That is, the sampling location distribution is a product of the population location distribution and the sampling rate. The sampling location distribution for all sampling designs and populations can be defined analogously. The sampling designs for each population are given in Table \ref{Tab:SD}.

\begin{table}[h!]
   \centering
   \begin{tabular}{lclc} 
      \hline
\multicolumn{2}{c}{Population 1}   & \multicolumn{2}{c}{Population 2}  \\
Sampling rate & Pref./Inf. &  Sampling rate & Pref./Inf.\\
      \hline
 (a) $p_{1a}(\mathbf{s}) = k$   & N/N				& (a)  $p_{2a}(\mathbf{s}) = k$ & Y/N  \\
 (b)  $p_{1b}(\mathbf{s}) = {logit}^{-1}(\alpha_0 + \alpha_1 W(\mathbf{s}))$  & Y/Y &  (b) $p_{2b}(\mathbf{s}) = {logit}^{-1}(\alpha_0 + \alpha_1 W(\mathbf{s}))$  & Y/Y\\
 (c)  $p_{1c}(\mathbf{s}) \propto \frac{exp(\alpha_0 + \alpha_1 W(\mathbf{s}))}{\lambda_1(\mathbf{s})}$ & Y/Y  & (c)  $p_{2c}(\mathbf{s}) \propto \frac{1}{\lambda_2(\mathbf{s})}$ & N/Y \\
   \end{tabular}
   \caption{Sampling designs for each population. The designs that are preferential and/or informative are indicated by the Y (yes) and N (no). \label{Tab:SD}}
\end{table}



For some constant $k$ with $0 < k <1$, sampling design (a) for both populations is non-informative. 
For population 1, the sampling location distribution is independent of the variable of interest, $Z$, meaning it is non-preferential. 
For population 2, since $\lambda_2(\mathbf{s})$ and $Z(\mathbf{s})$ are dependent, and $\lambda_{2a} = k \lambda_2(\mathbf{s})$, this sampling is preferential. 
The specification of sampling design (b) induces dependence between the sampling distribution and $Z$ for both populations making it informative and preferential.
Lastly, sampling design (c) results in a sampling location distribution of $\lambda_{1c}(\mathbf{s}) \propto exp(\alpha_0 + \alpha_1 W(\mathbf{s}))$ and $\lambda_{2c}(\mathbf{s}) = k$ for some constant $k$. For population 1,  this sampling design is informative and preferential, yet results in the sampling location distribution to be independent of the population location distribution. For population 2, the sampling location distribution is non-preferential but the sampling design is informative.

\subsection{Variogram estimation}

For each sampling design, 100 independent realizations of sample data are obtained. Then, estimates of the exponential variogram parameters are obtained using the composite likelihood (CL) approach as well as both weighted versions of the composite likelihood. The first weighted version (WCL1) assigns weights using the sampling rates given in Table \ref{Tab:SD}. The second weighted version (WCL2) assigns weights using the sampling location distribution.

\begin{figure}[h]
\begin{center}
\includegraphics[scale=.65]{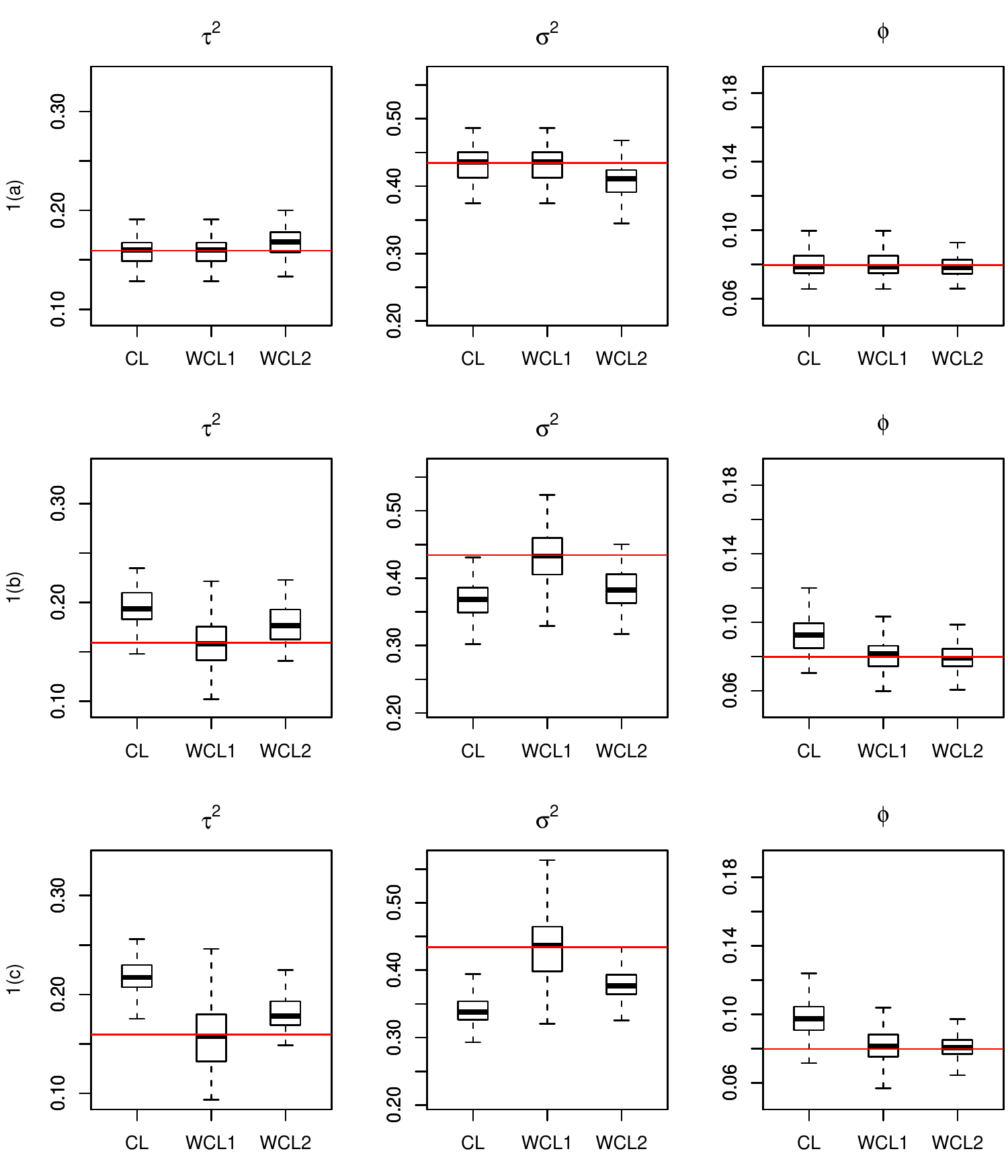}
\caption{Exponential variogram parameter estimates from population 1 under the three sampling designs. \label{Fig:Var1}}
\end{center}
\end{figure}

\begin{figure}[h]
\begin{center}
\includegraphics[scale=.65]{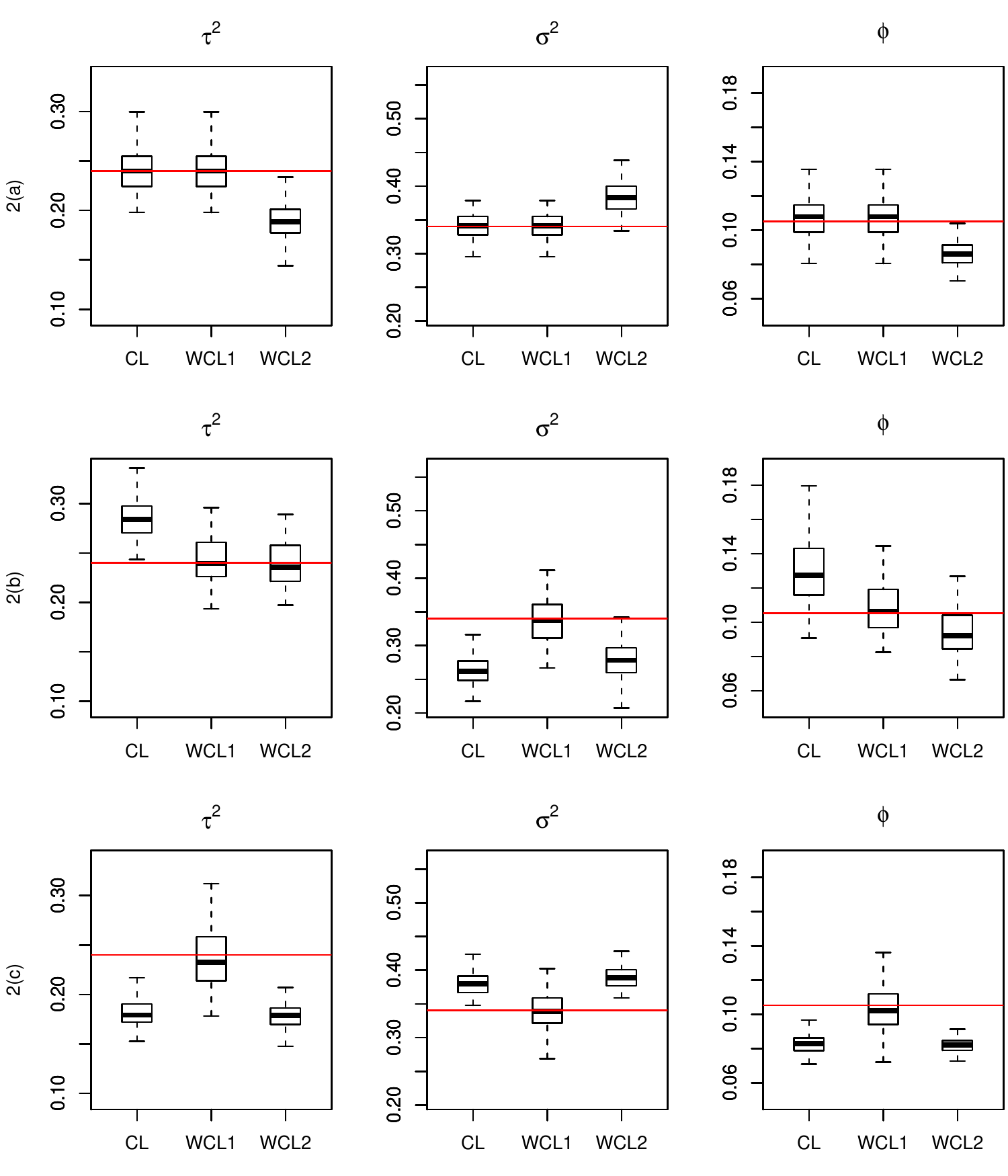}
\caption{Exponential variogram parameter estimates from population 2 under the three sampling designs. \label{Fig:Var2}}
\end{center}
\end{figure}

We compare the parameter estimates under each approach for each population and sampling design. 
Figures \ref{Fig:Var1} and \ref{Fig:Var2} show the distribution of the estimates of the exponential variogram parameters for both populations 1 and 2, respectively, obtained using CL, WCL1, and WCL2. In addition, the estimates obtained from the population are indicated by the horizontal line in each panel.  
For sampling design (a) that is non-informative, the variogram estimates obtained from the CL approach show very little bias relative to the population estimates. Recall that for population 2, sampling design (a) is preferential. 
The CL estimates obtained for sampling designs (b) and (c) are biased for both populations. These designs are informative for both populations but for population 2, sampling design (c) is non-preferential. 
The WCL1 estimates provide a reduction in bias for all three sampling designs, (a), (b), and (c) for both populations. 
The WCL2 estimates show improvement over the CL estimates for sampling designs that are both preferential and informative. 
On average, the WCL2 approach is less accurate than the WCL1 approach when compared to the population-based estimates.
For sampling design (a) for population 2 that is preferential and non-informative, the WCL2 estimates are more biased than the CL estimates. 
This is because, while the sampling location distribution and response variable are highly correlated, the sample is a simple random sample. 
Thus, the weights in (\ref{eq:wcl2}) induce bias in the parameter estimates.

Given the additional information on the sampling design, the simulation study suggests that the WCL1 approach is the preferred approach as it shows accurate spatial Gaussian process parameter estimation over all combinations of preferential and informative sampling designs considered.
The WCL2 approach is preferred over the CL approach only when the sampling design is known to be both preferential and informative.

\subsection{Kriging estimation}
Using the variogram parameter estimates for each of the 100 simulations obtained using CL, WCL1, and WCL2, we computed the kriging prediction means and variances for each population and sampling design.
In general, we detected little difference in the out-of-sample performance for the three parameter estimation approaches in terms of the generalized least squares estimator of $\mu$ or mean square prediction error. 
As anticipated, bias was highest under preferential and informative sampling designs and very little bias is detected under non-informative or non-preferential designs.
Recall that our primary focus is on kriging variance estimation under varying sampling designs. 
See \cite{Diggle2010}, \cite{Pati2011}, and \cite{Gelfand2012} for methods for reducing bias in kriging prediction estimates under preferential sampling.

%

\begin{figure}[h]
\begin{center}
\includegraphics[scale=.65]{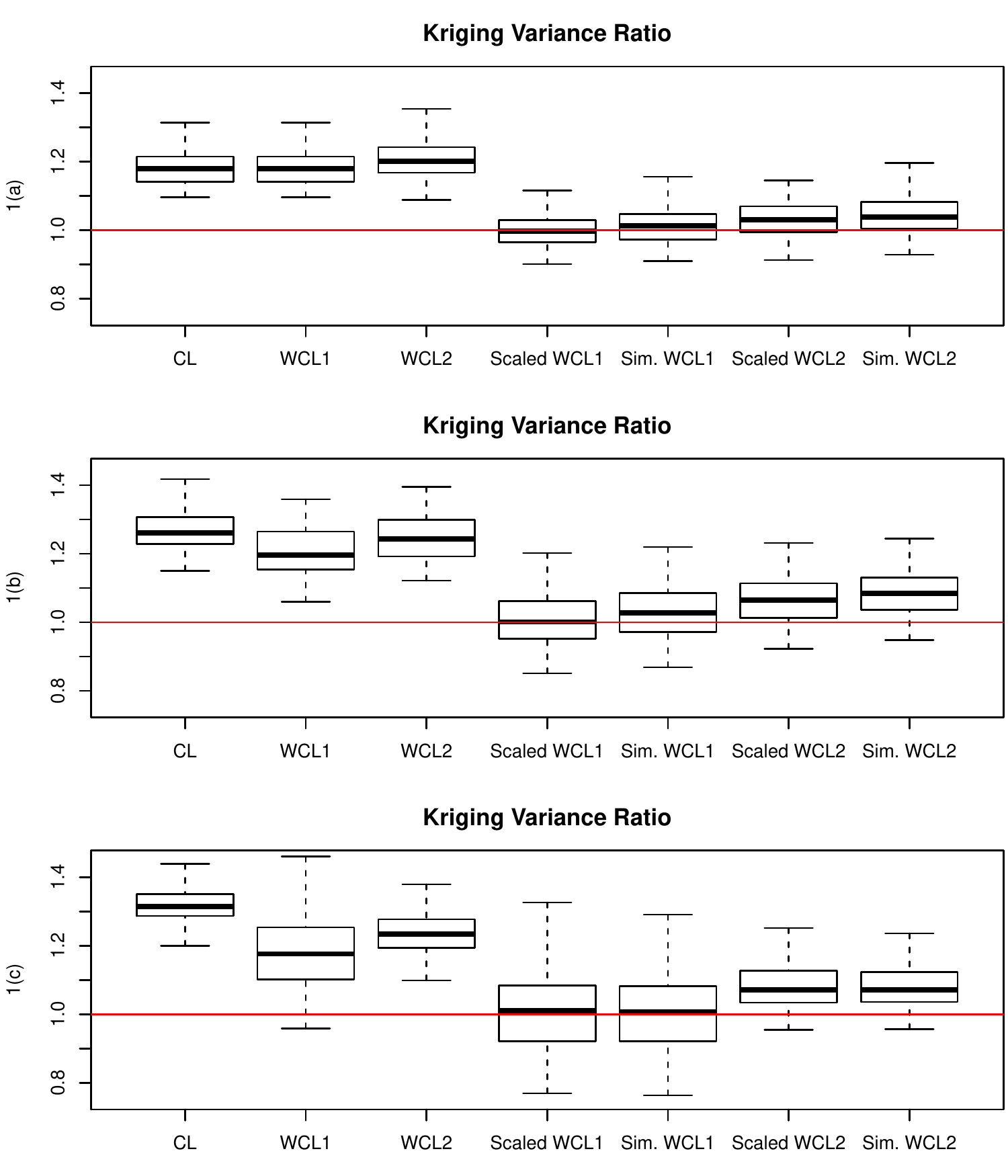}
\caption{Ordinary kriging variance ratio between the estimates obtained from the sample data and population data for each sampling design for population 1. Variance estimates were obtained using the CL parameter estimates, the WCL1 and WCL2 parameter estimates, and then using the WCL1 and WCL2 estimates and both the scaling and simulation approaches. \label{Fig:VarR1}}
\end{center}
\end{figure}
\begin{figure}[h]
\begin{center}
\includegraphics[scale=.65]{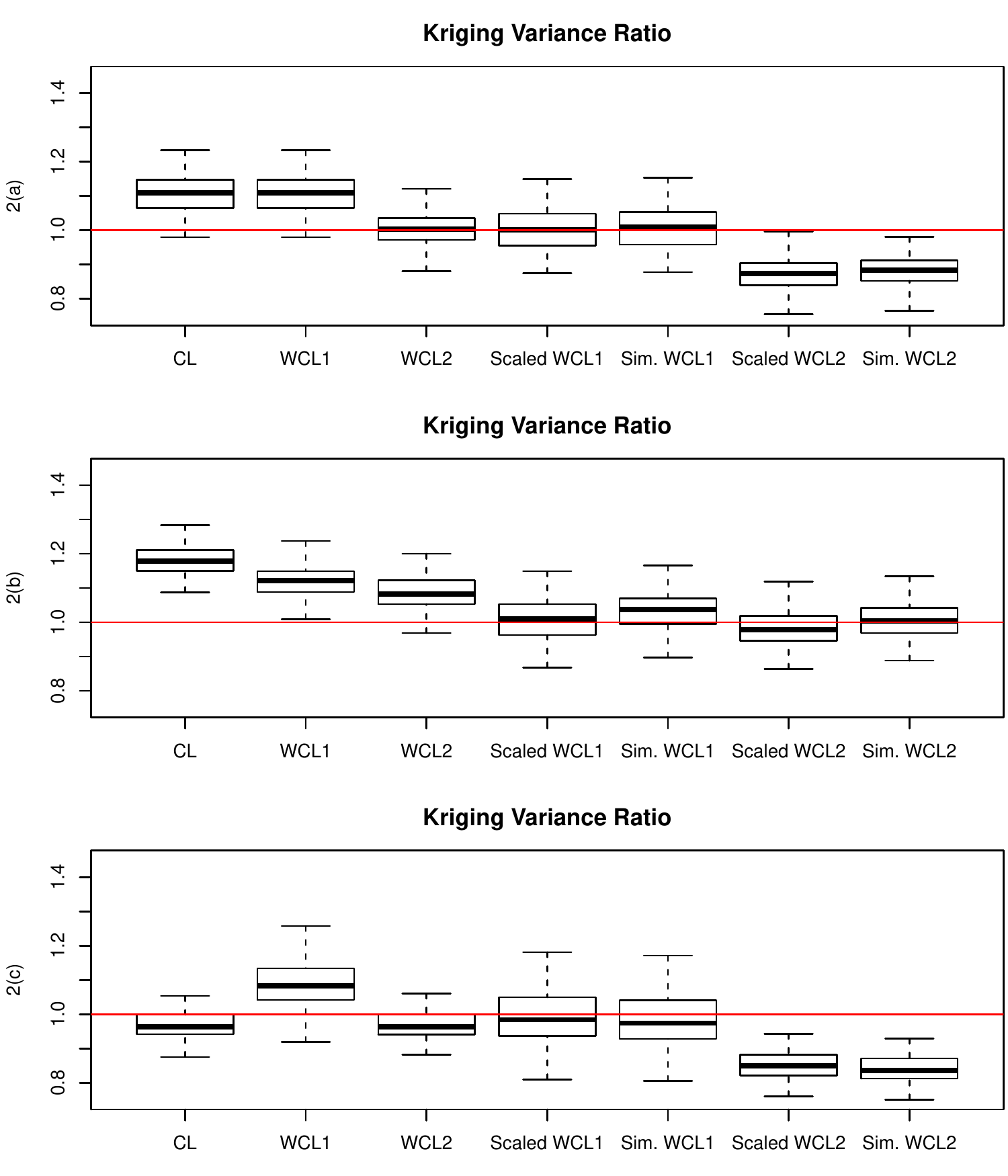}
\caption{Ordinary kriging variance ratio between the estimates obtained from the sample data and population data for each sampling design for population 2. Variance estimates were obtained using the CL parameter estimates, the WCL1 and WCL2 parameter estimates, and then using the WCL1 and WCL2 estimates and both the scaling and simulation approaches.\label{Fig:VarR2}}
\end{center}
\end{figure}

Figures \ref{Fig:VarR1} and \ref{Fig:VarR2} present the distribution of the ratio of the kriging variances obtained when using the sample data and population data for each design and population. Here, we compare the variance ratio between the sample and population using the CL, WCL1, and WCL2 parameter estimates. 
The kriging variance estimates based on the WCL1 and WCL2 parameter estimates are equal to or better at than those based on the CL for almost all population and sampling design combinations. 
One exception is population 2 under sampling design (c) which was informative but non-preferential. 

Kriging variances were also obtained using the scaling and simulation approaches proposed in Section \ref{sec3} given the WCL1 and WCL2 parameter estimates.
For both populations and all sampling designs, the variance ratio is close to 1 when using either the scaling or simulation procedures and the WCL1 parameter estimates, illustrating that these approaches can be used to obtain the population kriging variance estimates from sample data.
Using the WCL2 parameter estimates, the scaling and simulation procedures underestimated the population kriging variance for sampling designs (a) and (c) for population 2.
Recall that these sampling designs were preferential and non-informative and informative and non-preferential, respectively, for which the WCL2 parameter estimates were biased. 
For all other populations and sampling designs, the scaling and simulation approaches resulted in estimates of the kriging variance similar to those obtained from the population data.

\FloatBarrier
\section{Nitrate concentration in wells in California}
\label{sec:Wells}
We apply our methods to investigate nitrate concentration in wells in California. Similar data were analyzed by \cite{Chan2020} in studying various one and two stage sampling designs. In our analyses, we obtained nitrate concentrations from approximately 7000 wells. 
These data were collected between 2011 and 2021. 
Like in the analysis of \cite{Chan2020}, the nitrate concentrations were log-transformed and modeled using an exponential with nugget covariance function. 

\begin{figure}[h]
\begin{center}
\includegraphics[scale=.65]{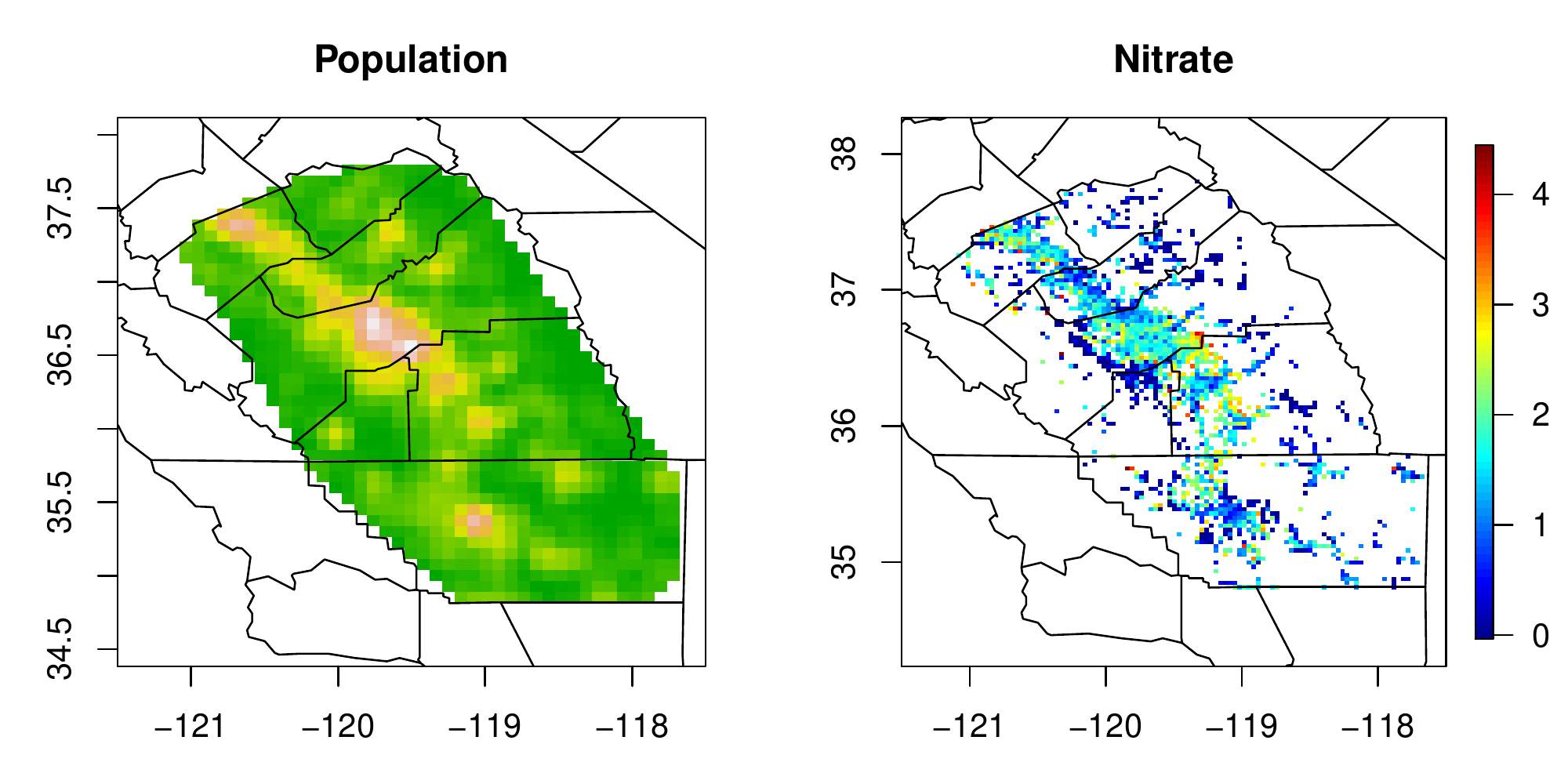}
\caption{(left) Kernel density estimate of the well locations across the region. (right) Log nitrate concentrations across the region. \label{Fig:Well1}}
\end{center}
\end{figure}

Figure \ref{Fig:Well1} shows a kernel density estimate of the well locations across the region. There are clear areas of high concentrations of wells in the central part of the region and spanning the region from the northwest to the southeast.
Also shown in Figure \ref{Fig:Well1} are the log nitrate concentrations for the region.
High concentrations in nitrate are found in the central part of the region. In general, these concentration levels appear positively correlated with the intensity surface (e.g., high concentration of nitrate in areas having high density of wells). 


We consider two different sampling scenarios to represent plausible sampling schemes employed by those collecting nitrate samples throughout the region. The first is a simple random sample (SRS). The second is an informative sampling design using stratification  based on the 69 sub-counties making up the region.  
The SRS assumed a sampling rate of 0.21. 
For the stratified sampling, the sampling rates within each sub-county were assumed constant and were specified inversely proportional to the number of wells in the sub-county. Specifically, let $N(\mathbf{s}_i)$ denote the number of wells in the sub-county containing location $\mathbf{s}_i$. Then, $p(\mathbf{s}_i) = 0.10$ for $N(\mathbf{s}_i)\geq 400$, $p(\mathbf{s}_i) = 0.15$ for $200 \leq N(\mathbf{s}_i) < 400$, $p(\mathbf{s}_i) = 0.20$ for $100 \leq N(\mathbf{s}_i) < 200$, $p(\mathbf{s}_i)= 0.30$ for $20 \leq N(\mathbf{s}_i) < 100$, $p(\mathbf{s}_i) = 0.40$ for $N(\mathbf{s}_i) <20$. These sampling rates were chosen such that the expected number of points were approximately equal for each sampling design.
For each sampling scenario, the sampling rate for each observation in the sample was assumed known. 
One realization from each of these sampling designs was generated for this analysis.
The random number of locations for the two sampling designs were 1432 and 1409, respectively.


Variogram estimates were obtained using the population data. 
In addition, for each sampling design, variogram estimates were obtained using the CL approach and the preferred WCL1 approach. For each design, the variogram estimates using the WCL1 approach are closer to those obtained using the population than those obtained by the CL approach.

\begin{table}[h!]
   \centering
   \begin{tabular}{lccc} 
      \hline
    & $\tau^2$ & $\sigma^2$ & $\phi$ \\
 \hline
 Population Estimates & 0.572 & 0.523 & 23.30\\   
      \hline
 CL\\        
 ~~~SRS & 0.576 & 0.534& 24.60\\
 ~~~Stratified & 0.601 & 0.516& 20.85\\
WCL1\\  
 ~~~SRS & 0.576 & 0.534& 24.60\\
 ~~~Stratified    & 0.569 & 0.523 & 24.59\\       
 \hline
   \end{tabular}
   \caption{Variogram parameter estimates obtained for the two sampling designs using both CL and WCL1. Also given are the parameter estimates obtained using the population data. }
   \label{tab:VarEsts}
\end{table}

\begin{figure}[h]
\begin{center}
\includegraphics[scale=.65]{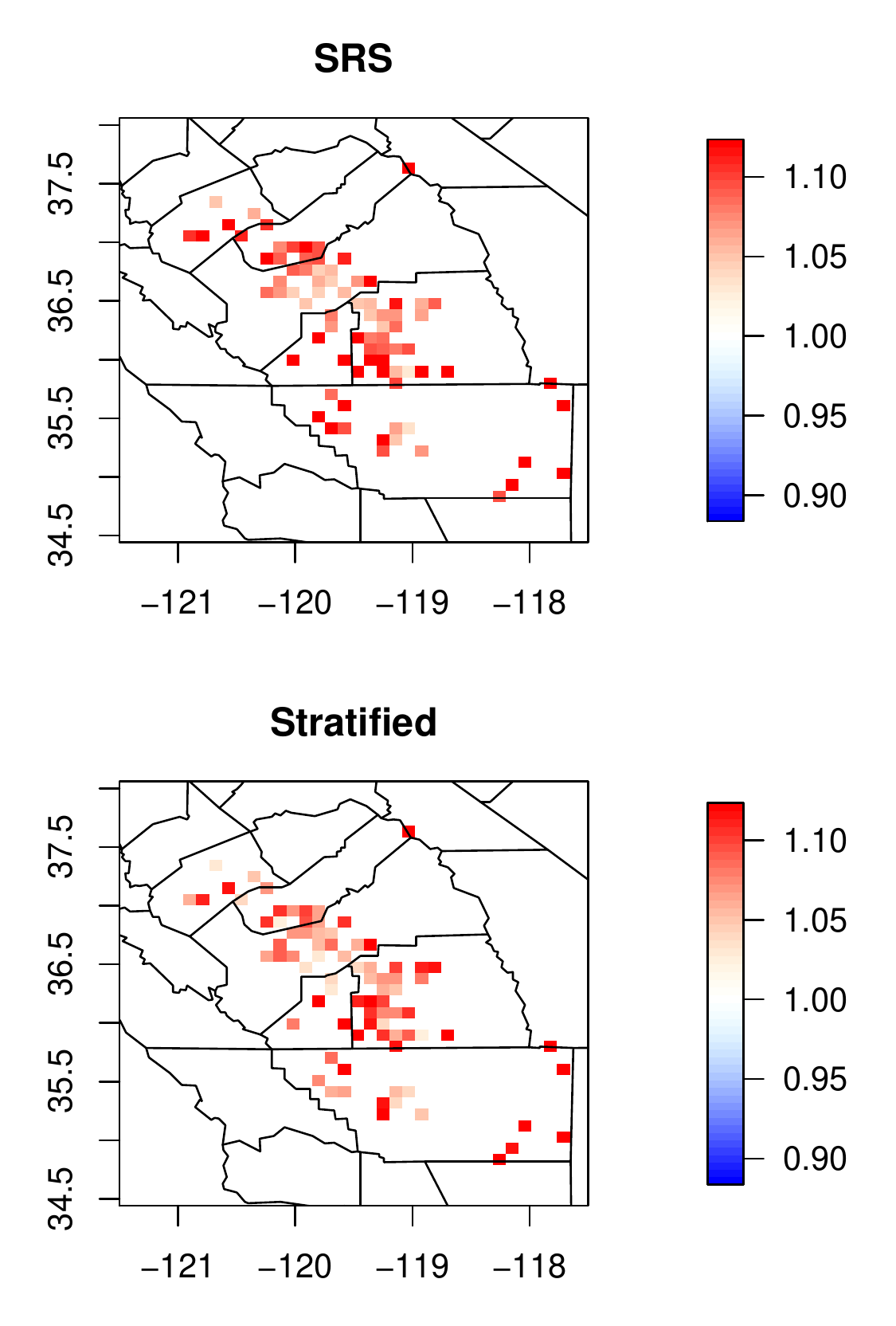}
\includegraphics[scale=.65]{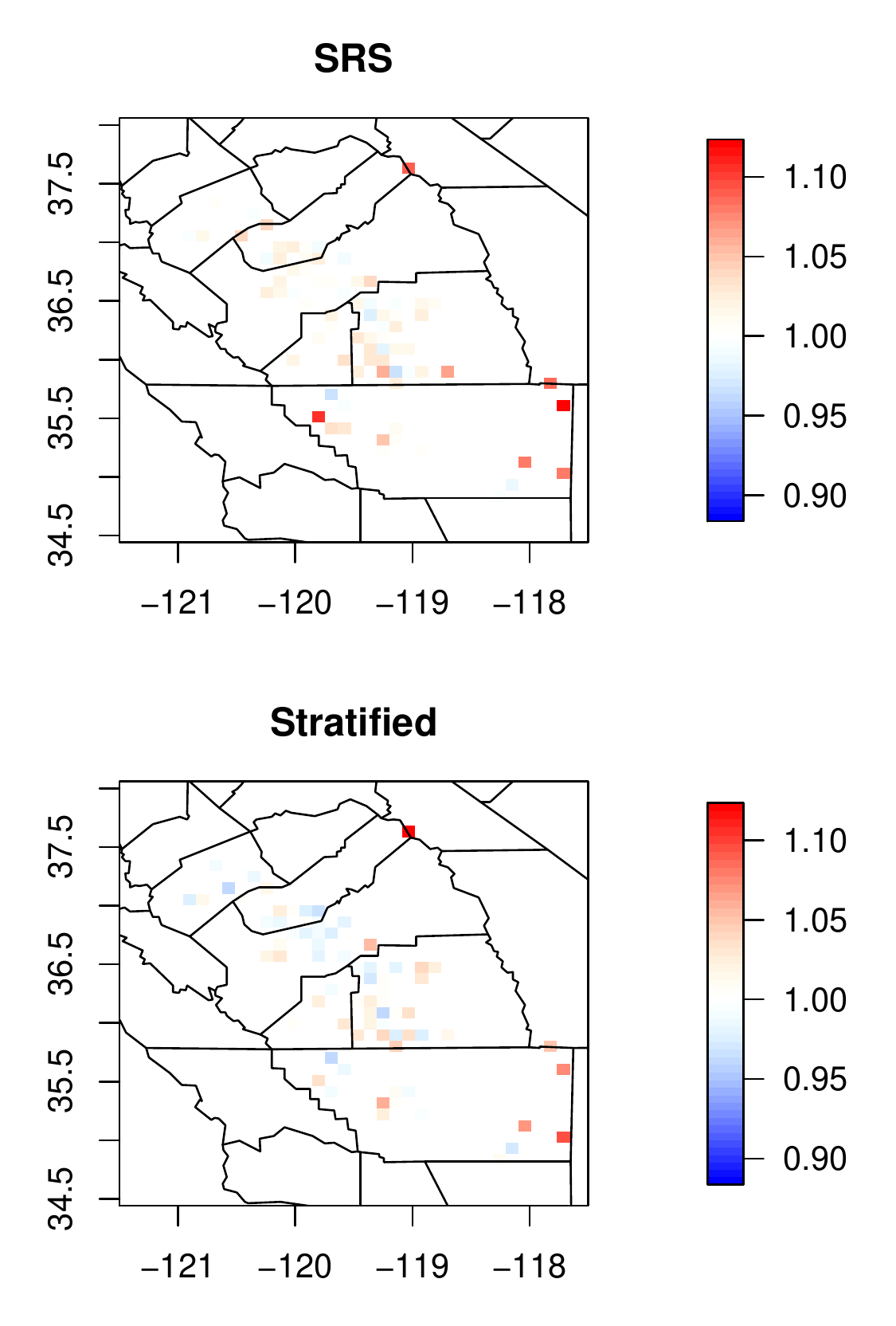}
\caption{Ordinary kriging variance ratio using the sample data relative to the population data for each of the sampling designs. The ratio is computed without the scaling (left) and with the scaling (right). \label{Fig:STD}}
\end{center}
\end{figure}

Using the variogram parameter estimates from the WCL1, we obtained kriging variance estimates for each sampling design for 100 out-of-sample well locations.
Variance estimates were obtained using the traditional kriging equations as well as using the scaling approach. 
The average kriging variance ratio using the sample data relative to the population data for the two sampling designs was 1.10 and 1.09, respectively. The ratio of the kriging variance estimates from the sample relative to the population are shown spatially in Figure \ref{Fig:STD} (left). Using the scaling approach, the average variance ratio for each sampling design was 1.02 and 1.01, respectively. Again, these are shown spatially in Figure \ref{Fig:STD} (right). The kriging variance estimates obtained using the scaling are very close to those obtained from the population (ratio $\approx$ 1) showing that this approach is able to estimate the kriging variances that one would have obtained had the entire population been sampled.

\section{Discussion}
\label{sec:Disc}

Sampling designs can impact spatial prediction by producing biased variogram estimates and/or kriging estimates. 
We proposed using a weighted composite likelihood approach that incorporates sampling designs when estimating variogram parameters.
These weights can be a function of either the known sampling rate or the estimated intensity function of the sample locations. 
Given the variogram estimates, we propose three approaches for estimating the population-based kriging variances using the sample data. The first approach assumes the locations of the unsampled data are known and can be used directly in defining the covariance matrices of the kriging equation.
When the locations of the unsampled locations are unknown, we developed the scaling and simulation approaches for kriging variance estimation.
Through simulation, we showed that the scaling and simulation approaches in combination with the weighted composite likelihood parameter estimates can mitigate the impacts of sampling design on kriging. 

The primary contribution of this work is the development of methods for correcting spatial Gaussian process parameter estimation under informative sampling designs. 
Under informative sampling designs, we found the weighted composite likelihood with weights a function of the sampling rates to be the best approach for covariance parameter estimation. 
This approach performed as well as or better than either the composite likelihood or sampling intensity weighted composite likelihood approach under any of the population and sampling design location and process distributions.
Whereas the sampling intensity weighted composite likelihood approach was not uniformly better than the unweighted approach, parameter estimation was improved when the sampling design was both informative and preferential. 
Therefore, in practice we recommend testing for preferential sampling \citep{Schlather2004, Guan2007, Watson2021} prior to using this approach. 
Under preferential and informative sampling, combining our approaches for correcting spatial parameter estimation with the joint modeling approaches \citep{Diggle2010, Pati2011, Gelfand2012} that account for preferential sampling for improved prediction and inference is a direction of future work. 
Since the kriging mean is a function of the spatial covariance parameters, we anticipate improved prediction when accounting for preferential sampling in estimating both the first and second order structure.

The scaling approach has computational advantages over the simulation approach.
Recall that the kriging equations involve matrix inversion.
The scaling approach results in a covariance matrix having the same dimension of the sample data, $m \times m$. 
Conversely, the simulation approach produces a covariance matrix of dimension $n \times n$, which may preclude this approach for large $n$.
The advantage of the simulation approach is that it is able to capture fine scale variation in the sampling process since the pseudo-observations are obtained as a function of $p(\mathbf{s})$, which can be highly variable across the region. 
On the other hand, the scaling approach assumes $p(\mathbf{s}^*)$ is constant near location $\mathbf{s}^*$. That is, a constant scaling, $p(\mathbf{s}^*)$ is applied to the covariance function between $Z(\mathbf{s}^*)$ and $Z(\mathbf{s}_i)$ for all $\mathbf{s}_i \in \mathcal{S}$.
An investigation into the possible bias in the scaling approach under varying specifications of the sampling rate process, $p(\mathbf{s})$, is the subject of future work. 



In our analysis, we focused on spatial kriging under informative sampling for finite population. 
We assumed throughout that the spatial process was stationary and that the sampling rate was known for the set of sample locations. 
Ongoing work includes investigating whether these results translate to non-stationary processes.
In addition, it is common in observational studies that one or both the sampling rates and population will be unknown.
With unknown sampling rate and population size, the sampling intensity weighted composite likelihood approach can still be used for variogram estimation. 
For a finite population, the simulation approach for kriging could be employed where the pseudo-observation locations would be retained only as a function of the sample location intensity. 
Future research is necessary to investigate the effects of informative and preferential sampling on spatial prediction for unknown populations.

\section*{Acknowledgement} The work of the authors was supported by the National Institute of Statistical Science.

\bibliographystyle{apalike}
\bibliography{PrefSamp}

\end{document}